\documentclass[aps, amssymb, amsmath, nofootinbib, showpacs, notitlepage, byrevtex, showkeys, prd]{revtex4-1}

\usepackage{graphicx}
\usepackage{bm}
\usepackage{bbm}
\usepackage{hyperref}
\usepackage{bbm}    
\usepackage{color}  
\usepackage[sc]{mathpazo}
\usepackage[T1]{fontenc}

\newcommand{\Id}{\mathbb{1}}

\begin{document}
	
\title{Branching of Center Vortices in $SU(3)$ Lattice Gauge Theory}
	
\author{Felix Spengler, Markus Quandt and Hugo Reinhardt}
\affiliation{Institut f\"ur Theoretische Physik\\
	Auf der Morgenstelle 14\\
	D-72076 T\"ubingen\\
	Germany}
\date{\today}
	
\begin{abstract}
   We analyze the branching of center vortices in $SU(3)$ Yang-Mills theory in maximal center gauge. 
   When properly normalized, we can define a branching probability that turns out to be 
   independent of the lattice spacing (in the limited scaling window studied here).
   The branching probability shows a rapid change at the deconfinement phase transition
   which is much more pronounced in space slices of the lattice as compared to time slices.
   Though not a strict order parameter (in the sense that it vanishes in one phase)
   the branching probability is thus found to be a reliable indicator for both the 
   location of the critical temperature and the geometric re-arrangement of vortex matter 
   across the deconfinement phase transition.
\end{abstract}
\maketitle
	
\section{Introduction} \label{Abschn: Einleitung}
The center vortex picture is one of the most intuitive and prolific explanation of 
colour confinement in strong interactions. It was first proposed by Mack and Petkova
\cite{mack}, but lay dormant until the advent of new gauge fixing techniques 
which permitted the detection of center vortex structures directly within
lattice Yang-Mills configurations \cite{DelDebbio}. These numerical studies have 
revealed a large mount of evidence in favour of a center vortex picture of confinement:
The center vortex density detected on the lattice in the maximual
center gauge after center projection properly scales with the lattice constant in the 
continuum limit and therefore must be considered a physical quantity \cite{Langfeld:1997jx}.
When center vortices are removed from the ensemble of gauge field configurations the 
string tension is lost in the temporal Wilson loop. Conversely, keeping the center vortex 
configurations only, the static quark potential extracted from the temporal Wilson loop 
is linearly rising at all distances \cite{DelDebbio}. Center vortices also seem to 
carry the non-trivial topological content of gauge fields: the Pontryagin index can 
be understood as self-intersection number of center vortex sheets in four Euclidean 
dimensions \cite{Engelhardt:1999xw,Reinhardt:2001kf} or in terms of the writhing number 
of their 3-dimensional projection which are loops \cite{Reinhardt:2001kf}. 
For the colour group $SU(2)$, attempts to restore the structure of the underlying (fat) vortices 
suggest that the topological charge also receives contributions from the colour structure of 
self-intersection regions of such fat vortices \cite{Nejad:2015aia,Nejad:2016fcl}.
Removing the center vortex content of the gauge fields makes the field configuration 
topological trivial and simultaneously restores chiral symmetry. The Pontryagin index
\cite{Bertle:2001xd} as well as the quark condensate \cite{Gattnar:2004gx,Hollwieser:2008tq} 
are both lost when center vortices are removed, see also 
\cite{Reinhardt:2003ku,*Reinhardt:2002cm}. In the case of $SU(3)$, 
this link of center vortices to both confinement and chiral symmetry breaking
has also been observed directly in lattice simulations of the low lying hadron 
spectrum \cite{OMalley:2012}.
Finally, the center vortex picture also gives 
a natural explanation of the deconfinement phase transition which appears as a 
depercolation transition from a confined phase of percolating vortices to a smoothly 
interacting gas of small vortices winding dominantly around  the compactified 
Euclidean time axis \cite{Engelhardt:1999fd}. 

Center vortices detected on the lattice after center projection form loops in $D = 3$ dimensions
and surfaces in $D = 4$; in both cases, they live on the \emph{dual} lattice and are closed 
due to Bianchi's identity. While a gas of closed loops can be treated analytically, see 
e.g.~\cite{Oxman:2017tel}, an ensemble of closed sheets is described by string theory, 
which has to be treated numerically. The main features of $D = 4$ center vortices detected 
on the lattice after center projection, such as the emergence of the string tension or 
the order of the deconfinement transition, can all be reproduced in an effective 
\emph{random center vortex model}: in this approach, vortices are described on a rather
coarse dual lattice (to account for the finite vortex thickness),  with the action given 
by the vortex area (Nambu-Goto term) plus a penalty for the curvature of the vortex sheets 
to account for vortex stiffness \cite{Engelhardt:1999wr,engelhardt,Quandt:2004gy}. 
The model was originally formulated for the gauge group $SU(2)$ \cite{Engelhardt:1999wr} 
and later extended to $SU(3)$ in Ref.~\cite{engelhardt}.

The $SU(3)$ group has two non-trivial center elements 
$z_{1/2} = e^{\pm i 2 \pi/3}$ which are related by $z^2_1 = z_2 \, , \, z^2_2 = z_1$. 
Due to this property two $z_1$ center vortices can fuse to a single $z_2$ vortex 
sheet and vice versa (see Fig.~\ref{fig:2} below). 
This vortex branching is a new element absent in the gauge group $SU(2)$. In 
Ref.~\cite{engelhardt} it was found within the random center vortex model that the 
deconfinement phase transition is accompanied with a strong reduction of the 
vortex branching and fusion. In the present paper we investigate the branching of 
center projected lattice vortices found in the maximal center gauge. 

This paper is organized as follows: In section \ref{sec:branch} we describe the
geometrical and physical properties of vortex branching and develop the 
necessary quantities to study this new phenomenon on the lattice. 
Section \ref{sec:setup} gives details on our numerical setup and the lattice 
parameters and techniques used in the simulations. The results are presented and 
discussed in section \ref{sec:results}, and we close with a short summary and an
outlook to future investigations. 

\section{Center vortex branching points}
\label{sec:branch}
On the lattice, center vortices are detected by first fixing all links $U_\mu(x)$ to 
a suitable center gauge, preferably the so-called \emph{maximal center gauge} (MCG),
cf.~eq.~(\ref{mcg}) below.
This condition attempts to find a gauge transformation which brings each link, on average, 
as close as possible to a center element. The transformed links are then projected on 
the nearest center-element, $U_\mu(x) \to Z_\mu(x) \in \mathbb{Z}_N$, and since it was 
already close, we can hope that the resulting $\mathbb{Z}_N$ theory preserves the relevant 
features of the original Yang-Mills theory. In fact, it has been shown that the 
string tension is retained to almost $100\%$ under center projection for the colour 
group $G=SU(2)$ and still to about $62\%$ for $G=SU(3)$ \cite{langfeld}, while the 
string tension disappears for all $G$ if vortices are removed \cite{DelDebbio, forcrand}. 
Also the near-zero modes of the Dirac operator relevant for chiral symmetry breaking disappear 
if vortices are removed from the physical ensemble \cite{Gattnar:2004gx,Hollwieser:2008tq}.

\begin{figure}[t]
	\begin{center}
	\includegraphics[width=4cm]{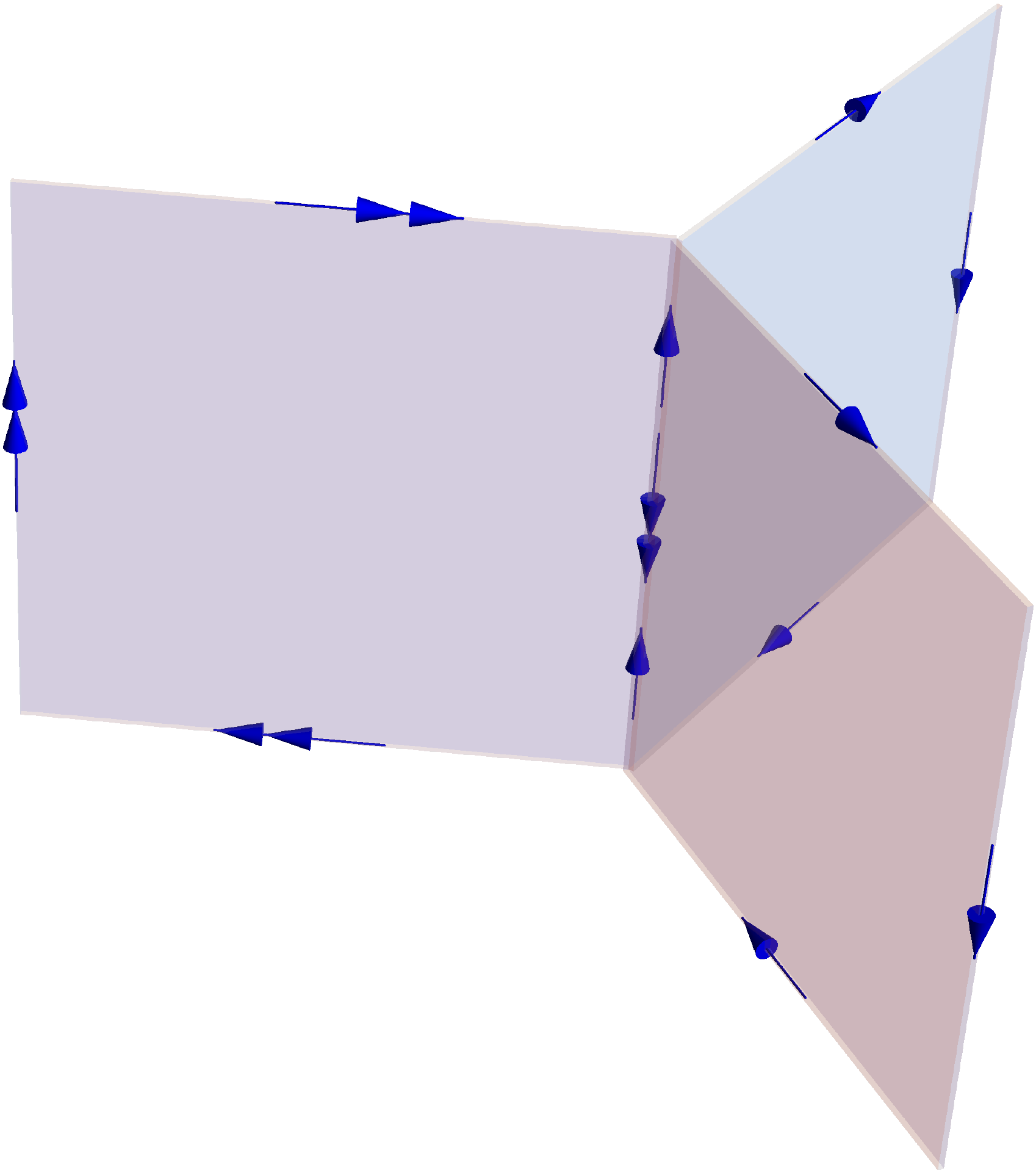}
	\hspace*{2cm}
	\includegraphics[width=4cm]{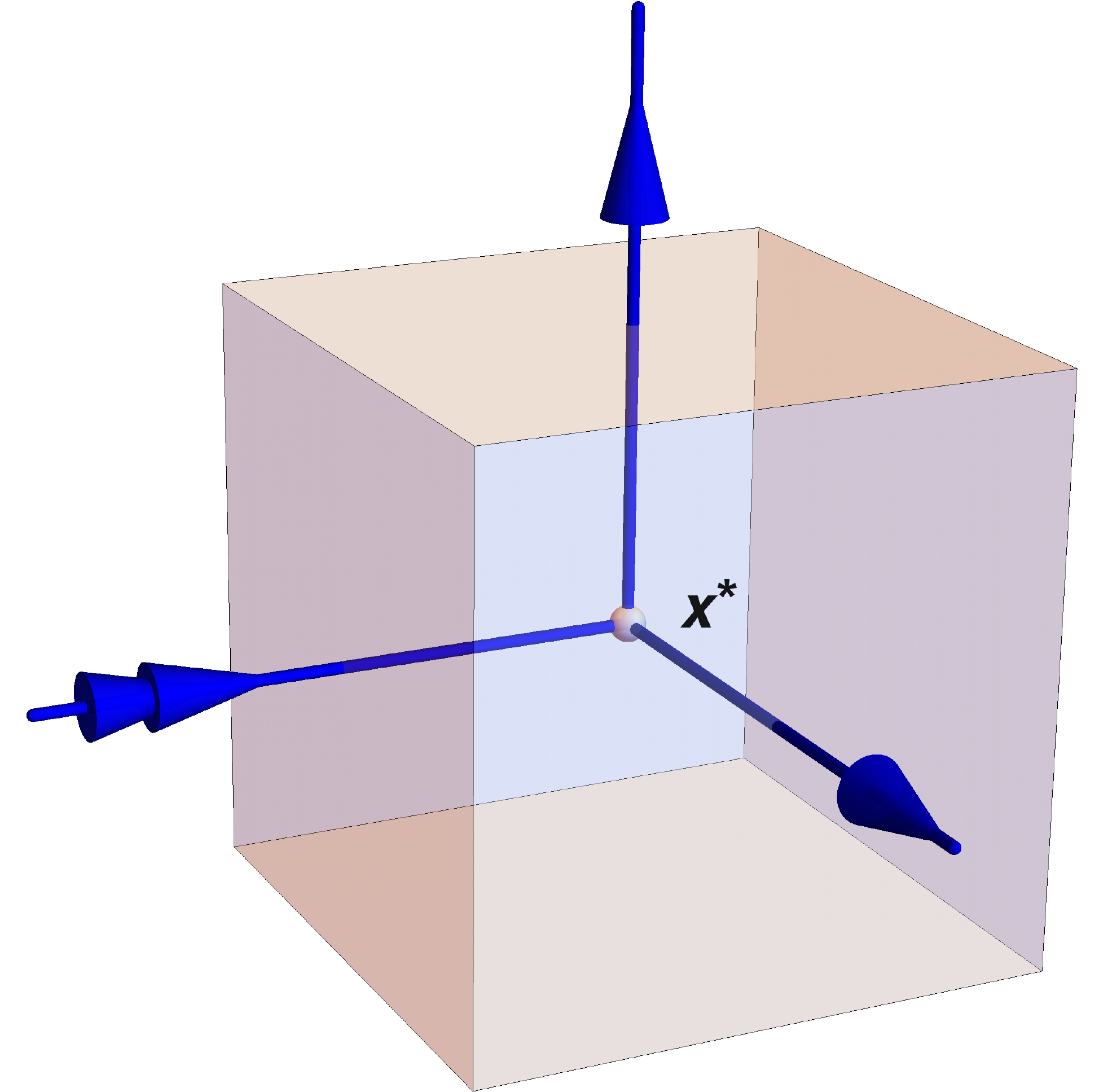}
	\hspace*{2cm}
	\includegraphics[width=3cm]{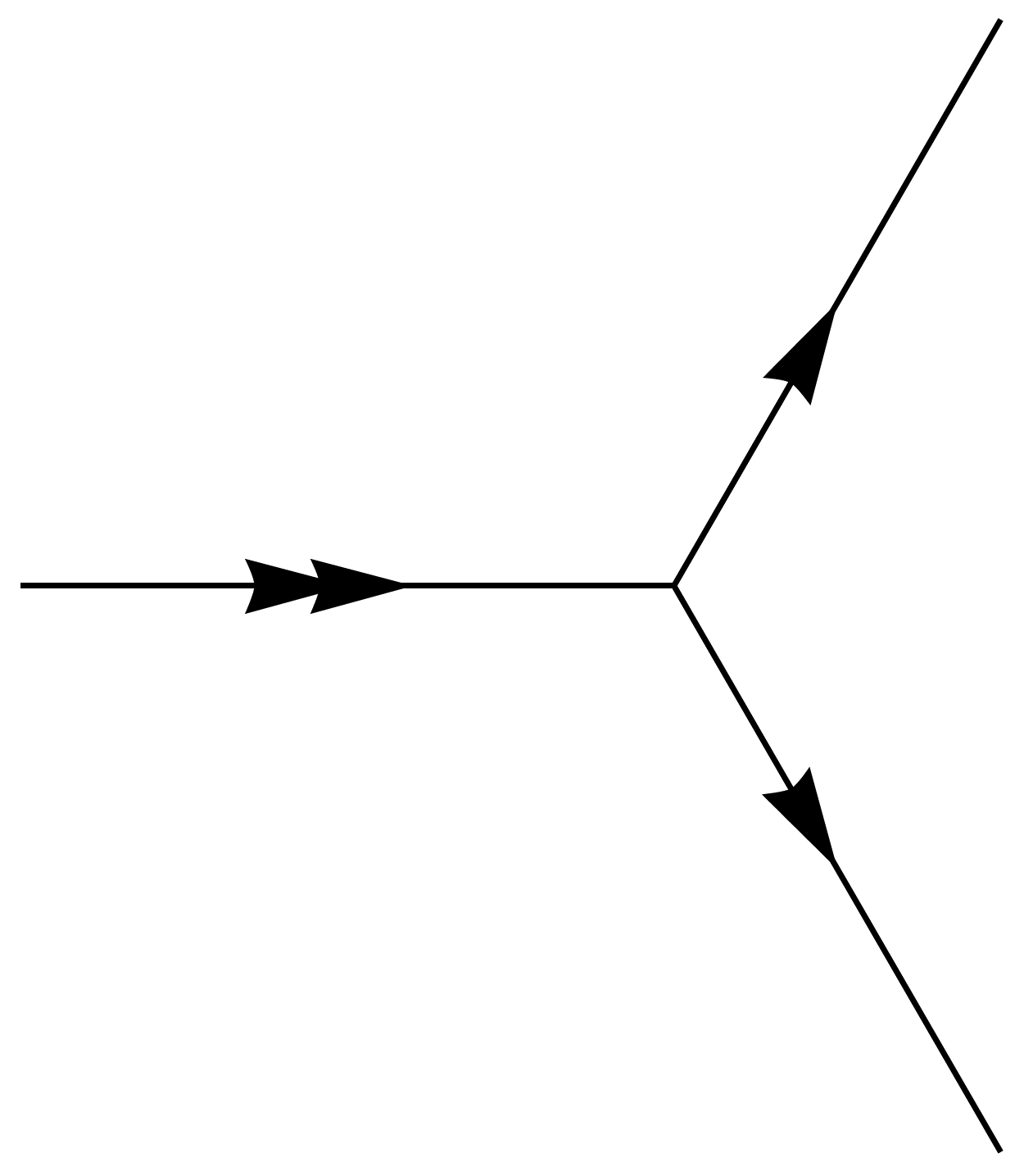}
	\end{center}
	\caption{Illustration of vortex branching. The single and double arrows on the lines 
	represent triality $q=1$ and $q=2$, respectively. The left figure represents a $\nu=3$ 
	vortex branching in the full $4D$ lattice. The graphic in the middle shows the same 
	situation from a 3D slice, where the vortex plaquettes are replaced by 
    three flux tubes joining at a branching point $x^\ast$. The tubes enter the elementary 
    cube surrounding $x^\ast$ by piercing three of its six surface plaquettes. The 
    right figure gives a simplified picture where only the branching vortex lines are displayed.}
	\label{fig:1}
\end{figure}

The center projected theory is much simpler to analyze. Since all links are center-valued
after projection, so are the plaquettes. If such a center-valued plaquette happens 
to be non-trivial, it is said to be pierced by a center vortex, i.e.~the corresponding 
\emph{dual} plaquette is considered part of a center vortex world sheet. For $G=SU(3)$, 
in particular, we associate a center projected plaquette $Z_{\mu\nu}(x)$ 
in the original lattice with a \emph{triality} $q_{\alpha\beta}(x^\ast) \in \{0,1,2\}$ 
on the dual lattice, where
\begin{align}
Z_{\mu\nu}(x) = \exp\left[i\,\frac{\pi}{3}\,\epsilon_{\mu\nu\alpha\beta}\,
q_{\alpha\beta}(x^\ast)\right]\,.
\label{triality}
\end{align}
Here, the usual sum convention over greek indices is in effect, and the footpoint of the dual 
plaquette is defined as $x^\ast = x + (\mathbf{e}_\mu + \mathbf{e}_\nu -
\mathbf{e}_\alpha-\mathbf{e}_\beta)/2$. As the reader may convince herself, this assignment 
is such that the initial and dual plaquette link with each other. The triality can be viewed
as a quantized flux of field strength flowing through the original plaquette. It is, 
however, only defined modulo $N=3$ so that a $q=1$ vortex is equivalent to $q=-2$, which 
in turn is a $q=2$ vortex with opposite direction of flux. This ambiguity gives rise to different
geometrical interpretations (see.~Fig.~\ref{fig:2}), but it does not affect the quantities 
studied in the present work. The vortex world sheet itself is now composed of all 
connected non-trivial dual plaquettes. This world sheet may \emph{branch} along 
links of the dual lattice where three or more vortex plaquettes join, cf.~the left 
panel of Fig.~\ref{fig:1}. 

For the actual measurement, we study the branching in the orignal 
lattice, where the branching link is dual to an elementary cube, while the plaquettes
attached to the branching link are dual to the plaquettes on the surface of the cube.
Geometrically, this can be visualized in the 3D slice\footnote{Such slices are obtained 
by holding either the Euclidean time coordinate $x_0$ (\emph{time slice}) or a 
space coordinate $x_i$ (\emph{space slice}) fixed.} of the original lattice which 
contains the cube, cf.~Fig.~\ref{fig:1}: in this slice, the vortex plaquettes 
are projected onto links which are dual to the non-trivial plaquettes and represent the 
center flux through the plaquettes. Vortex matter thus appears as a network of closed lines
composed of non-trivial dual links. These thin lines are the projection vortices in which 
the center flux of the unprojected (thick) vortex is compressed into a narrow tube with 
a cross section of only a single plaquette.

Vortex branching in a 3D slice occurs at \emph{branching points} which 
are the projection of the branching links in the 4D 
lattice. 
Geometrically, the branching points are located in the middle of the cubes dual to 
the branching links as illustrated in Fig.~\ref{fig:1}: the vortex lines 
entering an elementary cube must pierce the plaquettes on its surface,
and so up to six vortices can join at any given point of the dual 3D 
slice.\footnote{Equivalently, up to six vortex plaquettes in $D=4$ can join a common branching link.} 
We call this number $\nu(x^\ast) \in \{0,\ldots,6\}$ of vortex lines joining at a site 
$x^\ast$ of the dual 3D slice its \emph{branching genus}. Clearly, $\nu=0$ means that 
no vortex passes through $x^\ast$, while $\nu=2$ means that a vortex goes in and out 
without branching (but possibly changing its direction). 
The cases $\nu=4$ and $\nu=6$ correspond to vortex self-intersections (or osculation 
points), which are also present in the case of $G=SU(2)$.
The odd numbers $\nu=3$ and $\nu=5$, however, are genuine vortex branchings
which cannot be observed in $SU(2)$ and are thus a new feature 
of the center projected theory for the more complex colour group $SU(3)$.  
In the present study, we investigate the distribution of branching points in 
3D slices across the deconfinement phase transition.

It should also be mentioned that the case $\nu=1$ would represent a vortex end-point
which is forbidden by Bianchi's identity, i.e.~flux conservation modulo 3. More precisely,
Bianchi's identity in the present case states that the sum of the trialities of 
all plaquettes in an elementary cube of a 3D slice must vanish modulo $N$ (the number of 
colours). This holds even for cubes on the edge of the lattice if periodic boundary
conditions are employed. Clearly, this rule is violated if the cube has only $\nu=1$ 
non-trivial plaquette, which is hence forbidden. In our numerical study, the number of
$\nu=1$ branching points must then be exactly zero, which is a good test on our 
algorithmical book-keeping. 

Finally, we must also stress that $\nu=6$ branchings for the colour group $G=SU(2)$ 
are \emph{always} self-intersections or osculation points, while they can also be interpreted
as \emph{double vortex branchings} in the case of $G=SU(3)$. With the present technique, 
we cannot keep track of the orientation of vortices (i.e.~the direction of vortex flux), 
and hence are unable to distinguish double branchings from complex self-intersections. 
Fortunately, $\nu=6$ branching points are so extremely rare that they can be neglected 
entirely for our numerical analysis. If we speak of vortex branching, we thus always 
mean the cases $\nu=3$ and $\nu=5$, which only exist for $G=SU(3)$, and for which all
possible interpretations involve a single vortex branching.
Table \ref{tab:2} summarizes again the different sorts of 
vortex branchings and their geometrical meaning. 

\renewcommand{\arraystretch}{1.2} 
\begin{table}[h!]
	\centering
	\begin{tabular}{c|l}
		\toprule
		$\nu=0$ & no vortex\\
		$\nu=1$ & vortex endpoint, forbidden by Bianchi's identity\\
		$\nu=2$ & non-branching vortex\\
		$\nu=3$ & simple vortex branching\\
		$\nu=4$ & vortex self-intersection/osculation\\
		$\nu=5$ & complex vortex branching\\
		$\nu=6$ & complex vortex self-intersection/osculation/double branching
		\\ \botrule
	\end{tabular}
    \caption{Possible vortex branching types and their geometrical interpretation.
    As explained in Fig.~\ref{fig:2}, there is some arbitrariness in 
    the geometrical picture, while the branching genus $\nu$ is independent of all conventions.}
    \label{tab:2}
\end{table}

\section{Numerical setup}
\label{sec:setup}
We simulate $SU(3)$ Yang-Mills theory on a hypercubic lattice using the standard Wilson 
action as a sum over all plaquettes $U_P \equiv U_{\mu \nu}(x)$
\begin{align}
S=\sum_P \left[1-\frac{1}{2N}\,\mathrm{tr}(U_P + U_P^\dagger )\right].
\end{align}
Configurations are updated with the pseudo-heatbath algorithm due to Cabibbo and Marinari
\cite{su3heatbath} applied to a full set of $SU(2)$ subgroups. To study finite temperature, 
we reduce the extent $L_t$ of the Euclidean time direction, while keeping the spatial
extent $L_s \gg L_t$ to eliminate possible finite size effects,
\begin{align}
T = \frac{1}{a(\beta)\,L_t}\,.
\end{align}
Since the variation of $L_t$ only allows for a rather coarse temperature grid, we have 
also varied the lattice spacing $a(\beta)$ by considering three different couplings 
$\beta$ within the scaling window.\footnote{Finer temperature resolutions through the 
use of anisotropic lattices proved to be unnecessary for the present investigation.}
Table \ref{tab:1} lists the lattice extents and coupling constants used 
in our simulations. 

For each run, the lattice was thermalized using at least 100 heatbath sweeps, and 
measurements were then taken on $70$ to $200$ thermalized configurations (depending on $L_t$), 
with $10$ sweeps between measurements to reduce auto-correlations. For each measurement, 
the following sequence of steps was performed:

\begin{figure}[t]
	\begin{center}
		\includegraphics[width=2.5cm]{branch2b}
		\hspace*{2cm} 
		\includegraphics[width=2.5cm]{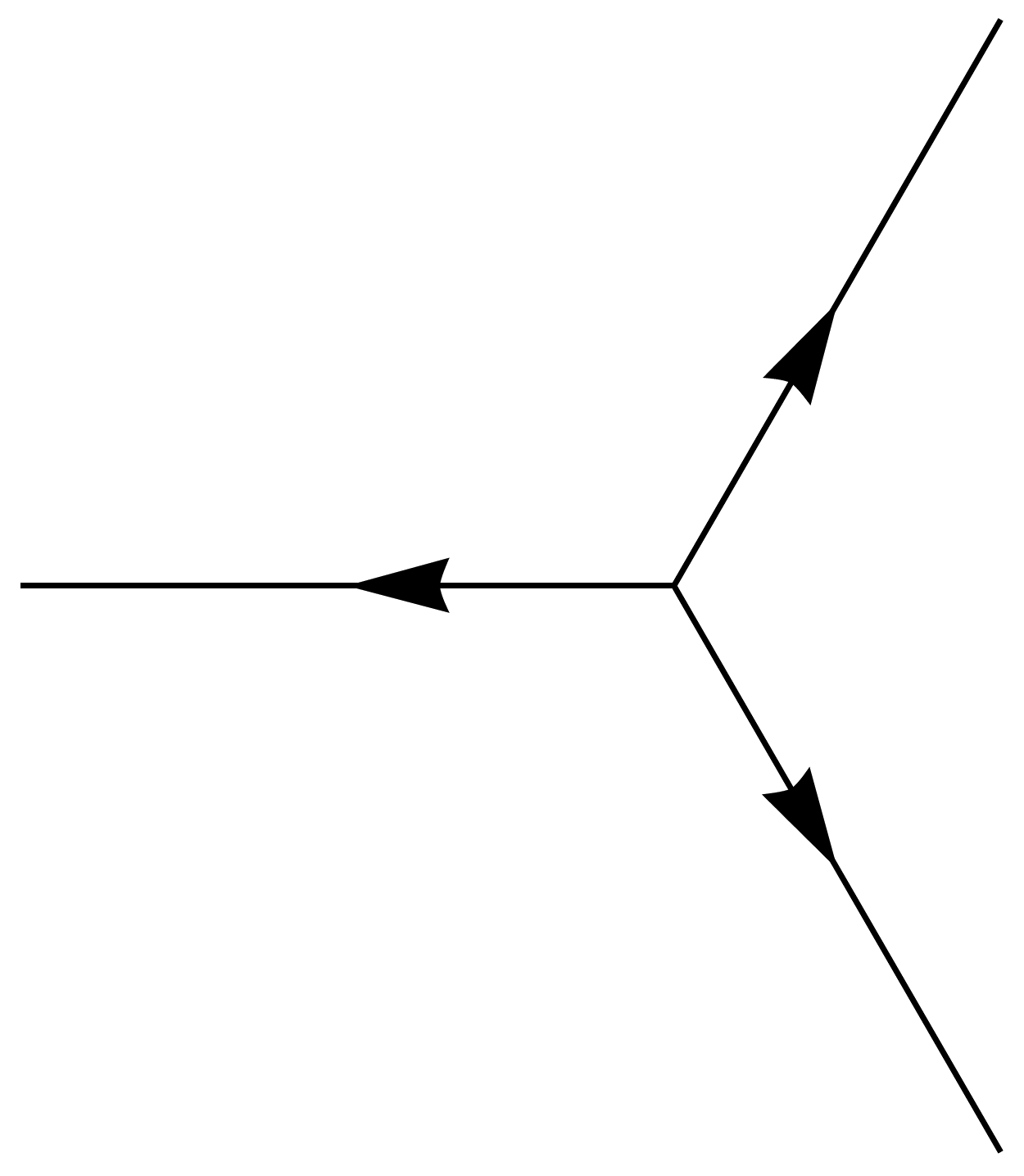}
		\\[2mm]
		\includegraphics[width=2.5cm]{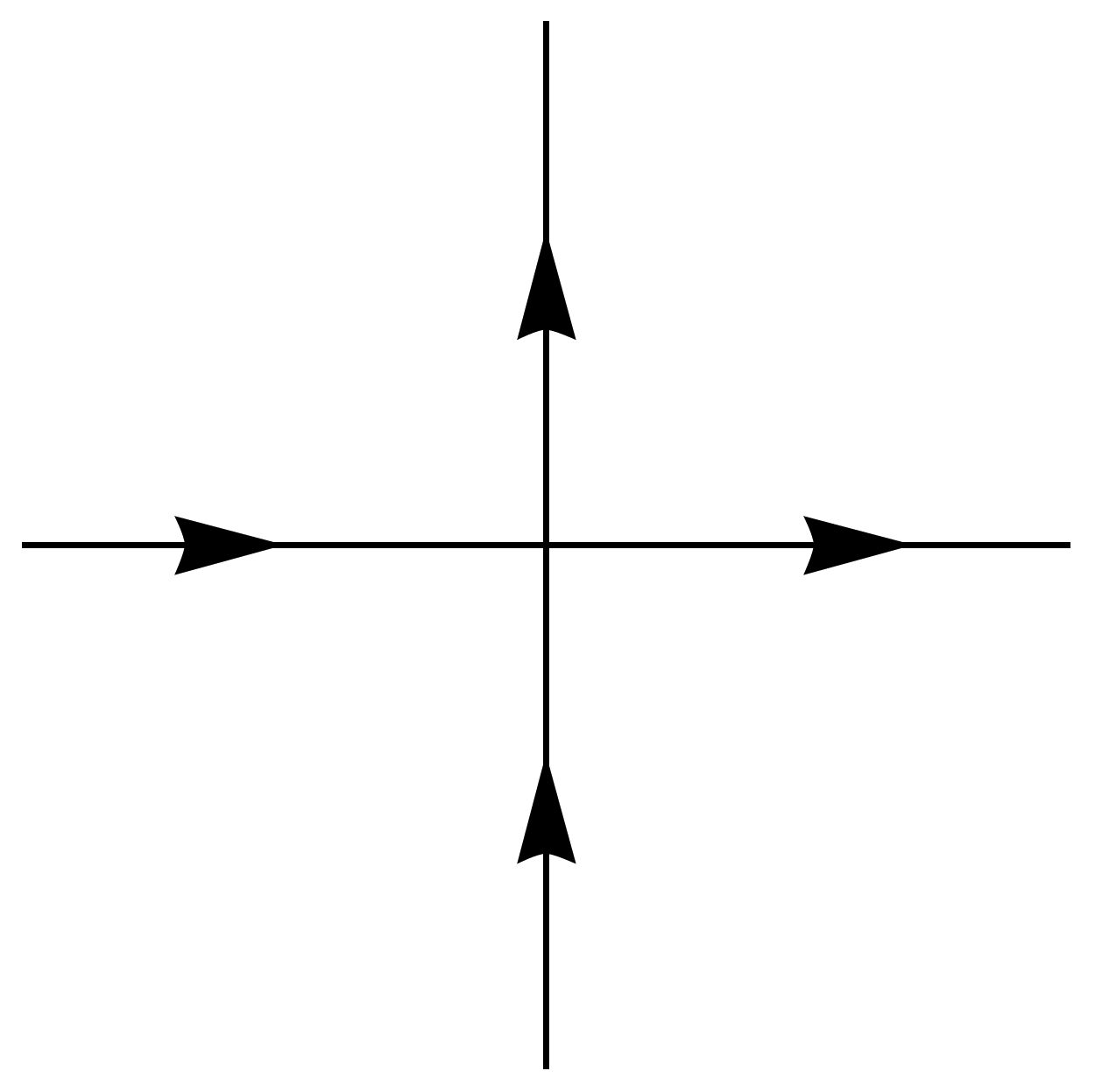}
        \hspace*{1cm}
        \includegraphics[width=2.5cm]{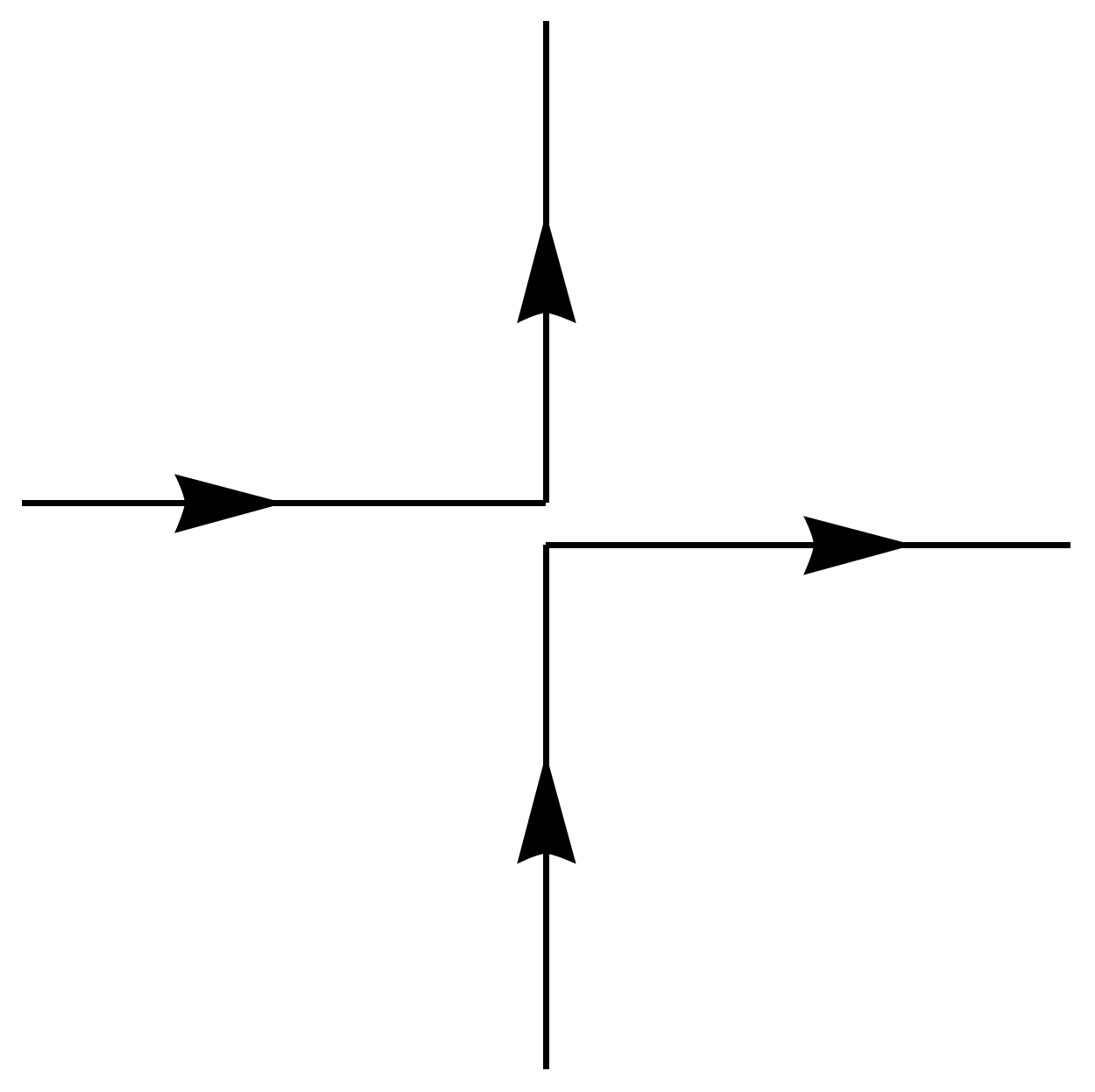}
        \hspace*{1cm}
        \includegraphics[width=2.5cm]{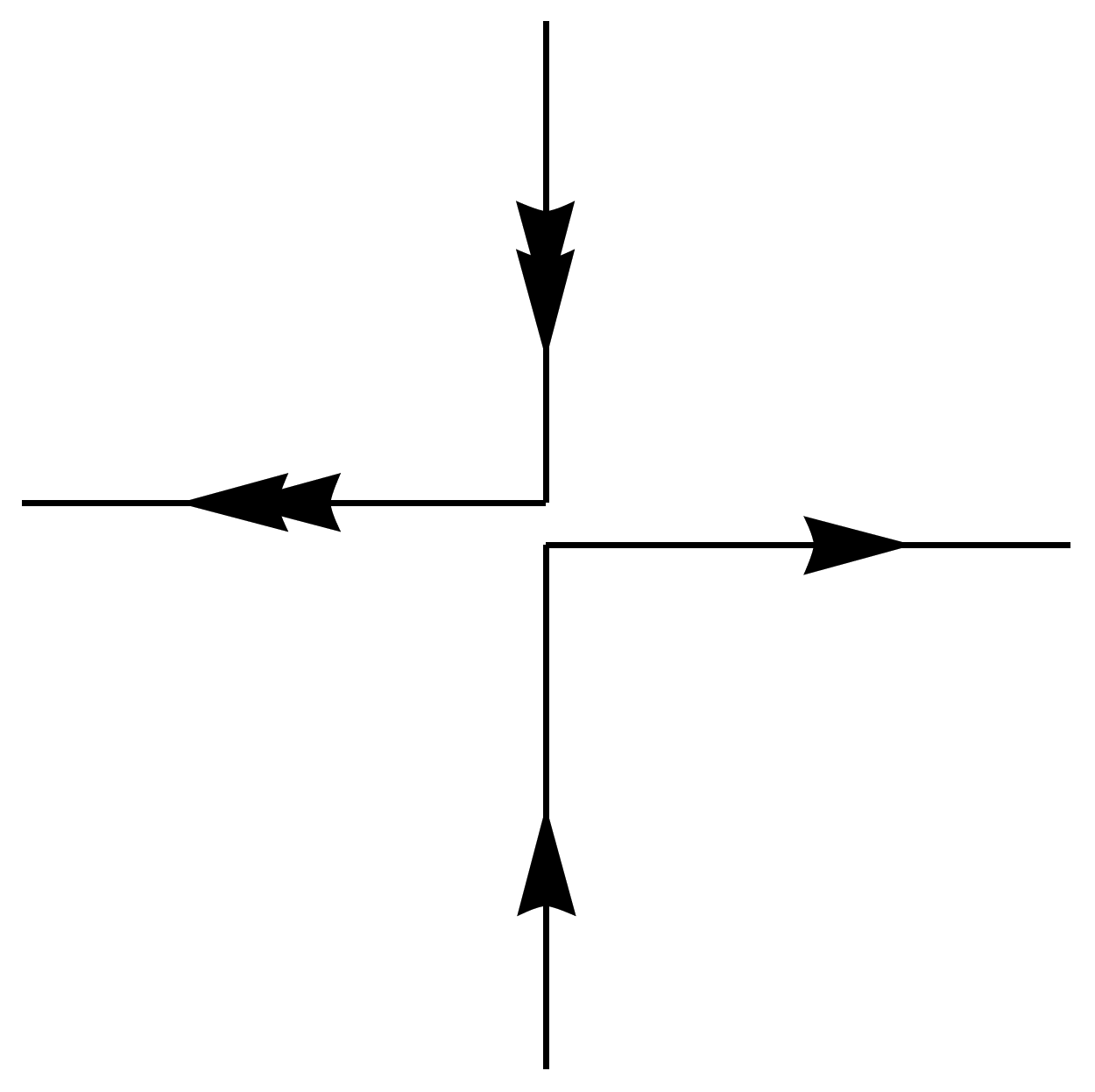}
	\end{center}
	\caption{Ambiguities in the interpretation of $SU(3)$ vortex branching.
		In the top line, the simple branching of a $q=2$ vortex on the left can be 
		equivalently described as three $q=1$ vortices emenating from a common source, 
		i.e. as $\mathbb{Z}_3$ center monopole. Similarly, the 
	    self-intersection of a $q=1$ vortex in the bottom line (left), is equivalent to 
        an osculation point of e.g. two $q=1$ vortices (middle) or a $q=1$ and a 
        $q=2$ vortex (right).}
	\label{fig:2}
\end{figure}

\medskip
\noindent\paragraph{Gauge fixing to maximal center gauge (MCG):} This is achieved by maximizing 
the functional
\begin{align}
F=\frac{1}{V} \sum\limits_{\{x,\mu\}} \left|\frac{1}{N}\,\mathrm{tr}\,U_\mu(x)\right|^2\,,
\label{mcg}
\end{align}
under gauge rotations, where $N=3$ is the number of colours and $V=\prod_\mu L_\mu$ is the 
lattice volume. The main gauge fixing algorithm used in this study is iterated 
overrelaxation \cite{overrelax} in which the local quantity
\begin{align}
F_x = \sum_\mu \left( \bigl|\mathrm{tr}\, \big\{\Omega (x) U_\mu (x)\big\} \bigr|^2 + 
\bigl|\mathrm{tr} \, \big\{U_\mu (x-\hat{\mu})\Omega^\dagger (x)\big\}  \bigr|^2\right)
\end{align}
is maximized with respect to a local gauge rotation $\Omega(x) \in SU(3) $ 
at each lattice site $x$. We stop this process when the largest relative change of $F_x$ 
at all sites $x$ falls below $10^{-6}$. More advanced g.f.~techniques such as 
simulated annealing \cite{gf_anneal} or Landau gauge preconditioners
\cite{gf_landau} from multiple random initial gauge copies have also been 
tested. While such methods are known to have a significant effect on the 
propagators of the theory in any gauge \cite{gf_green, gf_green2, *gf_green3}, 
we found that they have very little effect, at our lattice sizes, on the gauge 
fixing functional and the vortex geometry investigated here. For the production 
runs, we have therefore reverted to simple overrelaxation with random starts.

\medskip
\noindent\paragraph{Center projection:} Once a configuration is fixed to MCG, each link 
is projected to its closest center element $U_\mu(x) \rightarrow Z_\mu (x)$ by first 
splitting off the phase
\begin{equation}
\mathrm{tr}\, U_\mu(x) = \left|\mathrm{tr}\, U_\mu(x)\right| \cdot 
e^{ 2 \pi i \delta_\mu / N},
\end{equation}
which defines $\delta_\mu \in \mathbb{R}$ modulo $N$. After rounding $(\delta_\mu \,\mathrm{mod}\, N)$ 
to the closest integer $q_\mu \in [0,N-1]$, we can then extract the center projected link as 
\begin{align}
Z_\mu(x) \equiv \exp\left(i\,\frac{2\pi}{N}\,q_\mu\right)\Id \in \mathbb{Z}_N\,.
\end{align}
In the case of $SU(3)$, we will call the integer $q_\mu \in \{0,1,2\} $ the \emph{triality} of 
a center element. As mentioned earlier, the triality is only defined modulo 3, 
i.e.~$q_\mu= -2$ is identical to $q_\mu=1$. While this ambiguity alters the geometric interpretation
of a given vortex distribution (cf.~Fig.~\ref{fig:2}),
both the existence of a vortex branching point and its genus (the number of vortex lines meeting at 
the point) are independent of the triality assignment.  

\medskip
\noindent\paragraph{Vortex identification:} After center projection, all links are center valued, 
and so are the projected plaquettes. If such a center-valued plaquette happens 
to be non-trivial, we interpret this as a center vortex piercing the plaquette, i.e. the 
corresponding dual plaquette is part of the center vortex world sheet. The exact formula for 
the triality assignment of the vortex plaquettes was given in eq.~(\ref{triality}) above. For the 
computation of the area density of vortices, it is sufficient to consider a 2D plane in the 
original lattice and count the number of non-trivial plaquettes after center projection. 

\medskip
\noindent\paragraph{branching points:} As explained earlier, center vortices appear within a 
time or space slice as a network of links on the lattice dual to the slice. At each 
point $x^\ast$ of this dual 3D slice, between $\nu=0,2,\ldots,6$ vortex lines may join.
Since the point $x^\ast$ is the center of an elementary cube of the original time or 
space slice, the vortices joining in $x^\ast$ must enter or exit the cube and hence 
pierce some or all of the six plaquettes on its surface. We can thus determine 
$\nu(x^\ast)$ simply by counting the number of non-trivial plaquettes on elementary 
cubes in 3D slices of the lattice, and assign it to the possible branching point 
$x^\ast$ in the middle of the cube.

\renewcommand{\arraystretch}{1.2} 
\setlength{\tabcolsep}{8pt}
\begin{table}[t!]
	\centering
	\begin{tabular}{c||ccccc|ccccc|ccccc}
		\toprule
		$\beta$& \multicolumn{5}{c|}{$5.8$} & \multicolumn{5}{c|}{$ 5.85 $} & 
		\multicolumn{5}{c}{$ 5.9 $}\\
		$ L_t$ & 3 & 4 & 5 & 6 & 9 & 4 & 5 & 6 & 7 & 10 & 4 & 5 & 6 & 7 & 10 \\
		\# configs & 
		140 & 102 & 106 & 90 & 92 & 122 & 98 & 80 & 73 & 73 & 119 & 107 & 75 & 74 & 78
		\\ \botrule
	\end{tabular}
	\caption{\label{tab:1} Parameters for the finite temperature simulations. The last 
	row gives the number of configurations used for measurements, and the spatial 
	lattice size was $L_s = 24$ in all cases.}
\end{table}

\section{Results}
\label{sec:results}
The vortex area density is known to be a physical quantity in the sense that it scales 
properly with the lattice spacing $a(\beta)$ (see below) \cite{Langfeld:1997jx}. This entails that the overall 
amount of vortex matter quickly decays with increasing coupling $\beta$. To improve the 
statistics, we therefore choose coupling constants $\beta$ near the lower end of the scaling
window $ 5.7 \lesssim\beta \lesssim 7 $, cf.~table \ref{tab:1}. Since this implies a
rather coarse lattice, we must ensure that the lattice size in the short time direction 
does not become too small. For the values of $\beta$ chosen in our simulation, 
$L_t = \big[a(\beta)\,T\big]^{-1} \gg 1$ for temperatures at least up to $T\lesssim 2 T^\ast$, which is entirely 
sufficient for the present purpose. We have also checked that increasing the spatial volume 
from $L_s=16$ to $L_s=24$ has only marginal effects on the results, so that finite volume 
errors are also under control. In the final results, we only include the findings for the 
larger lattice extent $L_s = 24$.

\begin{figure}[t!]	
	\centering
	\includegraphics[width = 0.6 \textwidth]{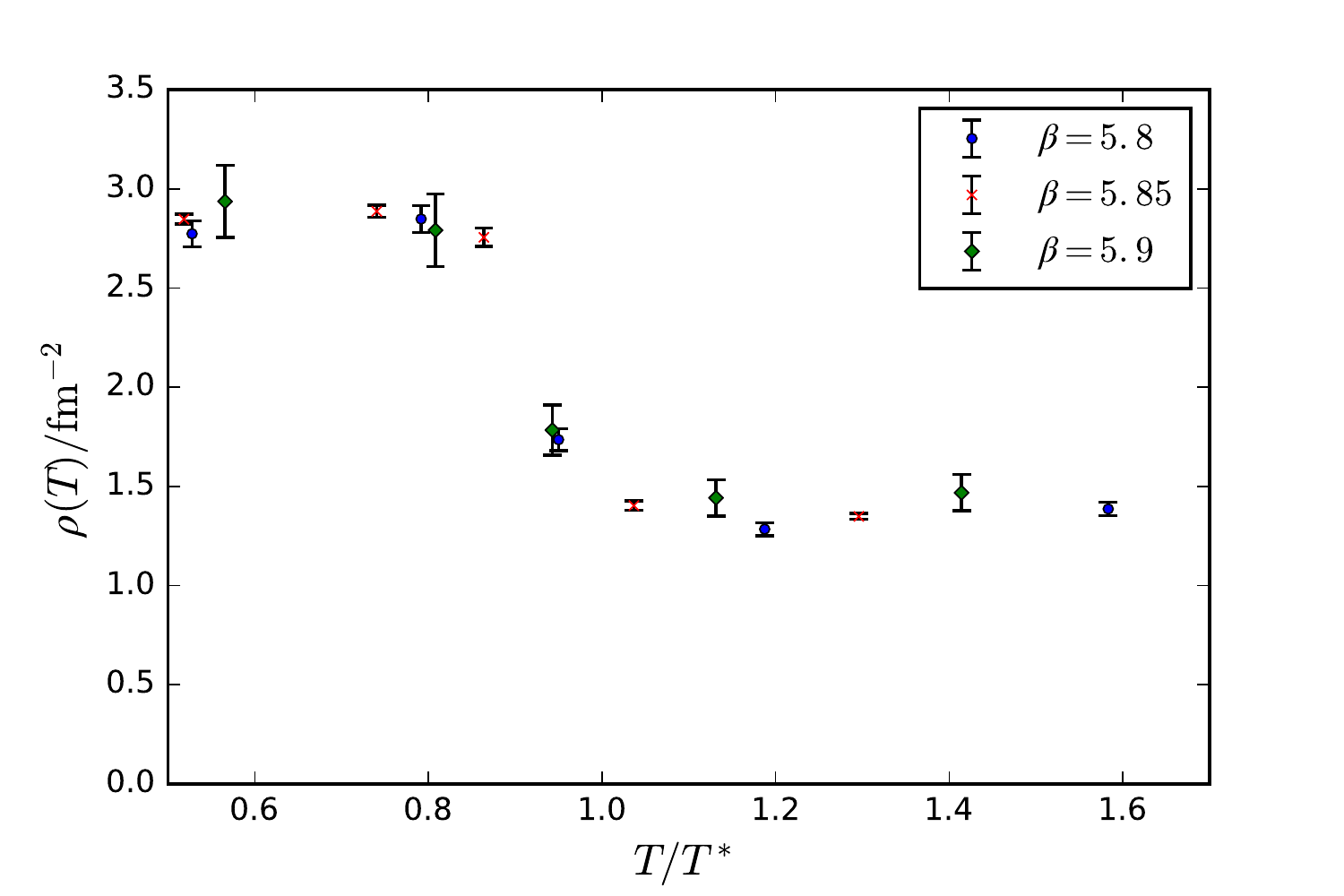}
	\caption{Vortex area density near the phase transition} 
\label{fig:3}
\end{figure}

The properties of vortex matter are intimately related to the choice and implementation 
of the gauge condition, as well as the absence of lattice artifacts. In particular, 
the vortex area density only survives the continuum limit if MCG is chosen and implemented 
accurately, and the lattice spacing is sufficiently small to suppress artifacts. 
As an independent test of these conditions, we have therefore re-analyzed the area 
density $\rho$ of vortex matter. In lattice units, this is defined as the ratio
\begin{align}
\hat{\rho}(\beta) = a(\beta)^2\,\rho = 
\frac{\# \text{non-trivial\, center plaquettes}}{\# \text{total\, plaquettes}}
\label{xvdens}
\end{align}   
in every 2D plane within the lattice. (We average over all planes in the full lattice 
or in appropriate 3D slices in order to improve the statistics.) After gauge fixing 
and center projection, the measurement of the vortex density is therefore a simple 
matter of counting non-trivial plaquettes. If we assume that the vortex area density
is a physical quantity that survives the continuum limit, we should have 
$\rho = c\,\sigma$, where $\sigma$ is the physical string tension and $c$ is a 
dimensionless numerical constant. A random vortex scenario \cite{mack} entails 
$\sigma = \frac{3}{2} \rho$ for $G=SU(3)$ which corresponds to $c=0.67$.
Previous lattice studies found a somewhat smaller value of about $c=0.5$ instead, 
indicating that the random vortex picture for MCG vortices at $T=0$ is not always 
justified \cite{langfeld}. In lattice units, these findings translate into
\begin{align}
\frac{\hat{\rho}(\beta)}{\hat{\sigma}(\beta)} = 
\frac{a(\beta)^2\,\rho}{a(\beta)^2\,\sigma} = \frac{\rho}{\sigma} = c \simeq 0.5
\qquad\qquad\text{indep.~of $\beta$ in scaling window}\,. 
\label{vdens}
\end{align}
For our values of the coupling as in table \ref{tab:1}, we have not measured the 
area density $\hat{\rho}(\beta)$ at $T=0$ directly, but instead took the data from the
largest temporal extent $L_t = 10$ which corresponds to a temperature 
$T/T^\ast \approx 0.55$ deep within the confined phase. Since the string tension and 
the vortex density do not change significantly until very close to the phase transition,
the $L_t=10$ data should still be indicative for the values at $T=0$. 
From these results and the string tension data $\hat{\sigma}(\beta)$ in Ref.~\cite{lucini}, 
the ratio (\ref{vdens}) can then be determined as follows:

\renewcommand{\arraystretch}{1.2} 
\setlength{\tabcolsep}{10pt}
\begin{table}[h]
	\centering
	\begin{tabular}{c|ccc}
		\toprule
		$\beta$& $5.8 $ &\ $ 5.85 $ &  $ 5.9 $\\
		$c$ & $0.558$ &  $0.573$  &  $0.591$ 
		\\ \botrule
	\end{tabular}
	\label{tab:3}
\end{table}

\noindent
As can be seen from this chart, the ratio (\ref{vdens}) is indeed roughly constant 
in the considered coupling range, and also in fair agreement with previous lattice 
studies \cite{langfeld}, given the fact that we did not really make a $T=0$ simulation. 
In addition, inadequately gauge fixed configurations would show increased randomness which 
would lead to a significant drop in the vortex density as compared to the string tension 
data from Ref.~\cite{lucini}, and hence a much smaller value of $c$.
We thus conclude that our chosen lattice setup and gf.~algorithm are sufficient for 
the present investigation.

\medskip
Next we study the finite temperature behaviour of vortex matter.
The critical deconfinement temperature for $G=SU(3)$ is given by
$T^\ast / \sqrt{\sigma} \approx 0.64$ \cite{lucini} . Since we do not measure the string tension 
independently, we can use eq.~(\ref{vdens})
\begin{align}
\frac{T^\ast}{\sqrt{\rho_0}} = \frac{T^\ast}{\sqrt{\sigma}}\,\sqrt{\frac{\sigma}{\rho_0}}
= \frac{T^\ast / \sqrt{\sigma}}{\sqrt{c}} \approx 0.90
\end{align}
to determine the critical temperature in units of the zero-temperature vortex density
$\rho_0 \equiv \rho(T=0)$ which sets the scale in our simulations. In absolute units,
\begin{align}
\sqrt{\rho_0} = \sqrt{c\,\sigma} \approx 330\,\mathrm{MeV}\,.
\end{align}
From the results in Fig.~\ref{fig:3} we see that there is roughly a 50\% drop
in the vortex density at the critical temperature, which is consistent with 
the findings of Ref.~\cite{langfeld}. A complete loss of 
vortex matter at $T^\ast$ would mean that both the temporal \emph{and} spatial string 
tension would vanish in the deconfined phase, contrary to lattice results \cite{sigma_spatial}. 
What happens instead is a \emph{percolation phase transition} in which the geometric 
arrangement of vortices changes from a mostly random ensemble to a configuration in 
which most vortices are aligned along the short time direction \cite{Engelhardt:1999fd}. 
Since this leads to a 
nearly vanishing vortex density in space slices while the average density only 
drops mildly, the density in time slices and the associated spatial string tension 
must even increase for $T > T^\ast$.  

This considerations imply that a good order parameter for confinement in the 
vortex picture should be sensitive to the randomness or order in the geometric
arrangement of vortex matter and, as a consequence, should behave differently 
in temporal or spatial 3D slices of the lattice. A prime candidate in 
$SU(3)$ Yang-Mills theory is the 3-volume density of \emph{branching points},
since it is directly defined in 3D slices and describes deviations of the 
vortex cluster from a straight aligned ensemble. This has previously been
studied in the effective center vortex model \cite{engelhardt} where indeed 
a significant drop of vortex branching was observed in the deconfined phase,
but not directly in lattice Yang-Mills theory.

\begin{figure}[t]
	\begin{center}
		\includegraphics[width=8.7cm]{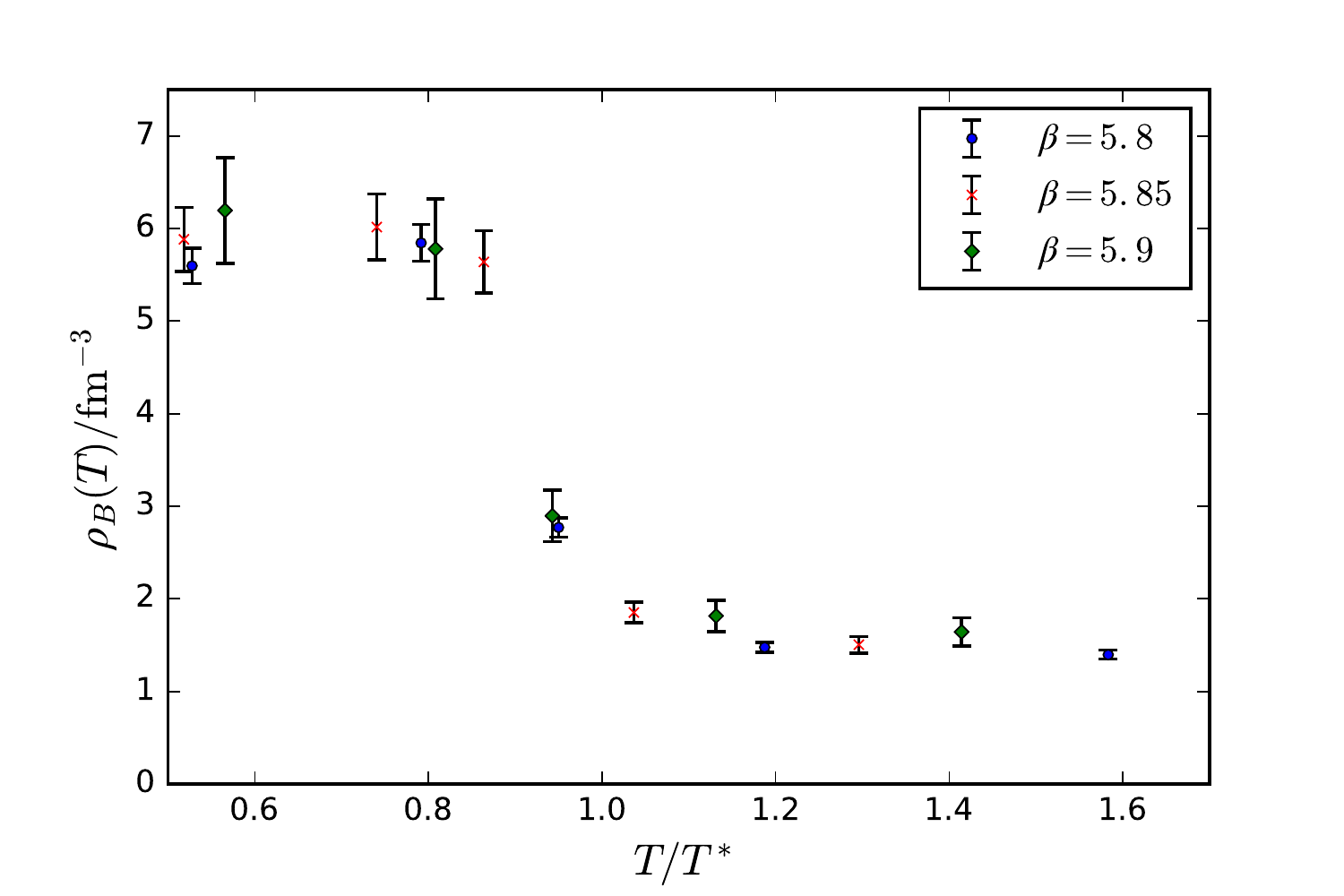}
		\includegraphics[width=8.7cm]{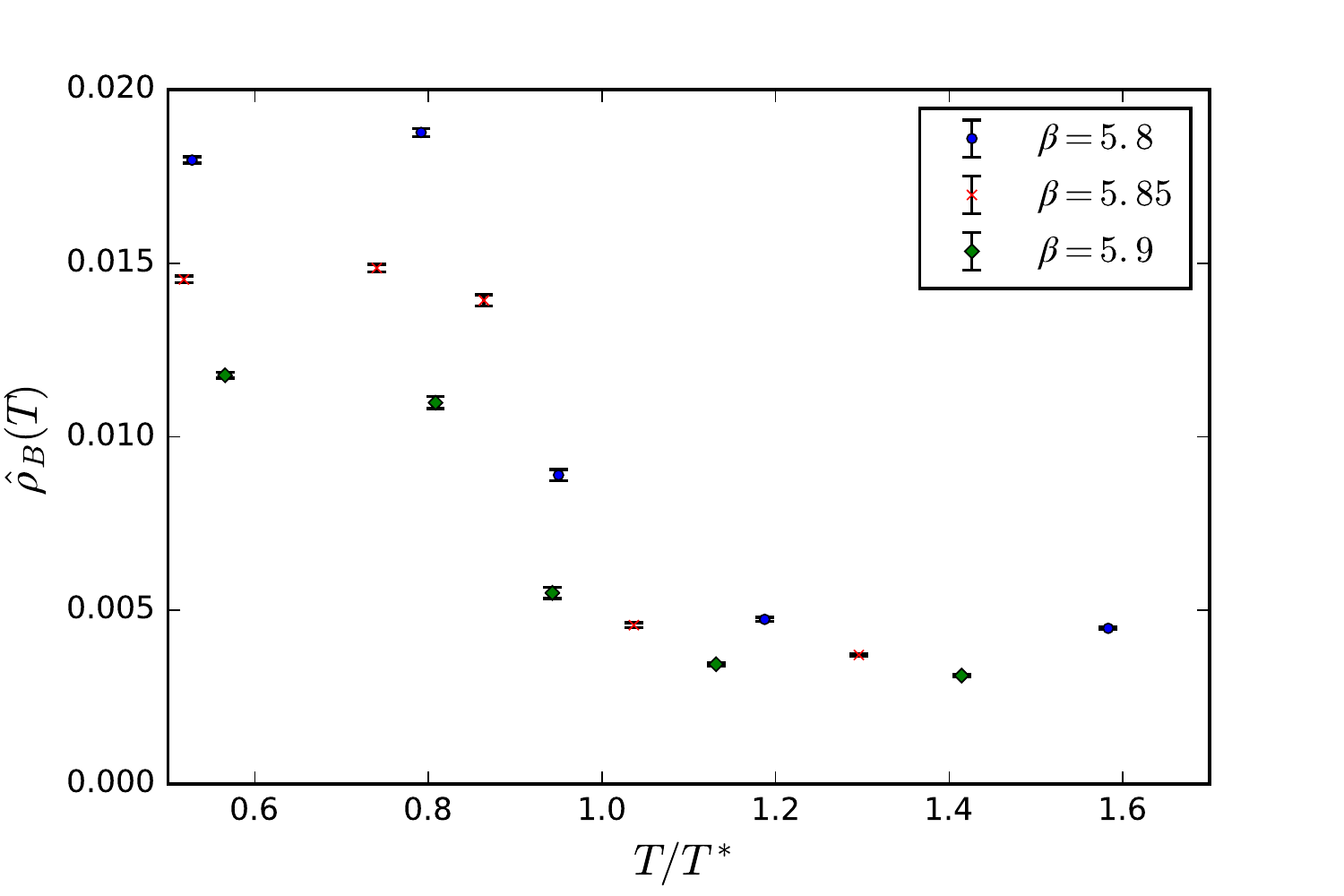}
	\end{center}
	\caption{Scaling of the volume density of vortex branching in space slices of the lattice.
		The physical density (\ref{rbx}) (\emph{left}) shows no apparent scaling violations.
		For comparison, the dimensionless density (\ref{rb}) (\emph{right}) shows 
		the amount of scaling violations to be expected for the present range of couplings.
		Error bars for the physical density are much larger since they also include 
		uncertainties in the physical scale taken from Ref.~\cite{lucini}.}
	\label{fig:4}
\end{figure}

Since vortex branching implies a deviation from a straight vortex flow, 
we expect that it is suppressed in the deconfined phase where most vortices
wind directly around the short time direction. In addition, the residual 
branching for $T > T^\ast$ should be predominantly in a space direction 
(since the vortices are already temporally aligned) and should hence be
mostly visible in \emph{time slices}, where the vortex matter is expected to 
still form large percolating clusters. In space slices, by contrast, 
vortices are mostly aligned (along the time axis) in the deconfined phase,
and the suppression of the remnant branching for $T > T^\ast$ should be 
much more pronounced.

\bigskip\noindent
To test these expectations, we have measured the (dimensionless)
volume density of branching points
\begin{align}
\hat{\rho}_B \equiv \frac{\text{\# branching points in lattice dual to 3D slice}}
{\text{\# total sites in lattice dual to 3D slice}} = 
\frac{\text{\# elementary cubes in 3D slice with $\nu \in \{3,5\}$ }}
{\text{\# all elementary cubes in 3D slice}}\,,
\label{rb}
\end{align}
by assigning the vortex genus $\nu \in \{0,\ldots,6\} $ to all elementary cubes 
in a 3D slice, cf.~section \ref{sec:branch}, and counting them. (To improve the 
statistics, we have averaged over space- and time slices separately using the 
same thermalized configurations.) Generally, we find 
\begin{enumerate}
	\item vortex endpoints with $\nu=1$ do not appear, i.e.~vortices are closed 
	in accordance with Bianchi's identity;
	\item vortex branchings are rare as compared to $\nu=2$ non-branching vortex matter;
	\item complex vortex branchings with $\nu=5$ are very rare and significantly 
	reduced as compared to the simple branchings with $\nu=3$; numerically, the 
	$\nu=5$ branchings contribute with only $0.1\ldots 1.0 \%$ to the total
	branching probability.
\end{enumerate}

To construct a quantity which has the chance of scaling to the continuum, we 
must express the branching density in physical units,
\begin{align}
\rho_B(T,\beta) \equiv \frac{\hat{\rho}_B(T,\beta)}{a(\beta)^3}\,,
\label{rbx}
\end{align}
where $a(\beta)$ is the lattice spacing at coupling $\beta$, which we take from 
Ref.~\cite{lucini}. Eq.~(\ref{rbx}) is 
indeed a physical quantity as can be be seen directly from the result in Fig.~\ref{fig:4}
where the data for all $\beta$ considered here fall on a common curve. Since we only 
considered a limited range of couplings $\beta$, one could be worried that 
possible scaling violations in $\rho_B$ would not be very pronounced. As can be seen from 
the right panel of Fig.~\ref{fig:4}, this is not the case: the dimensionless density (\ref{rb}), 
for instance, exhibits large scaling violations which are clearly visible even for our 
restricted range of couplings. This gives a strong indication that the branching density 
$\rho_B(T)$ really survives the continuum limit, even though further simulations at large 
couplings would be helpfull to corroborate this fact. 

\begin{figure}[t]
	\begin{center}
		\includegraphics[width=8.7cm]{rho_b_space}
		\includegraphics[width=8.7cm]{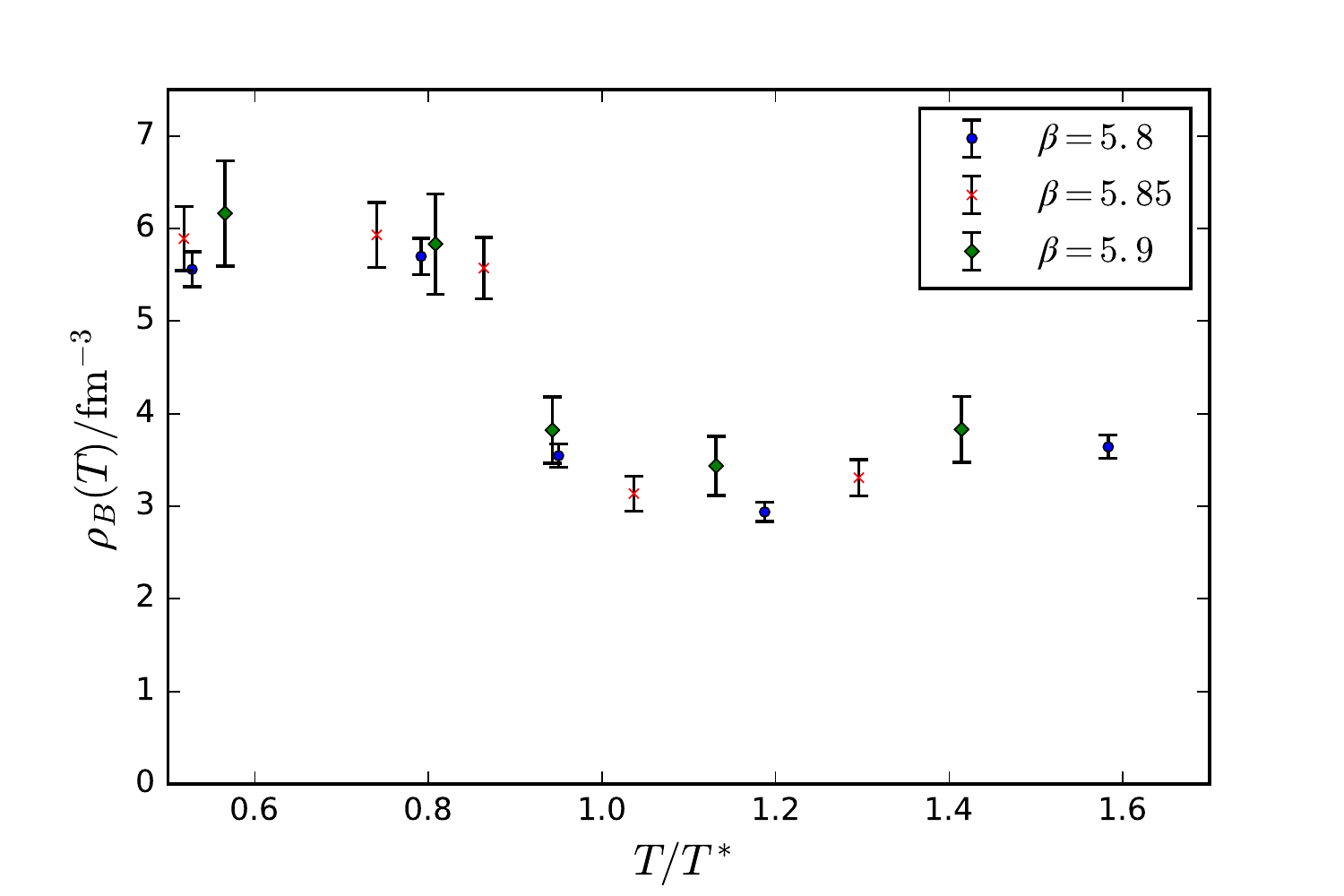}
	\end{center}
	\caption{Volume density of vortex branching points in physical units, measured in 
		space slices (\emph{left}) and time slices (\emph{right}).
		Error bars include statistical errors and uncertainties
		in in the physical scale taken from Ref.~\cite{lucini}.}
	\label{fig:5}
\end{figure}

From Fig.~\ref{fig:5}, the physical branching density indeed shows a
rapid drop at the critical temperature $T=T^\ast$, while it stays roughly 
constant below and above $T^\ast$. In particular, the maximal value is 
expected at $T \to 0$. We have not made independent measurements 
at $T=0$, but the available data from $L_t=9$ and $L_t=10$ corresponding to 
$T/T^\ast = 0.55$ should still be indicative for the value at zero temperature
since the vortex properties are known to show no significant change until very close 
to the phase transition. With this assumption, we find, in absolute units,
\begin{align}
\rho_B(0) \approx 5.86 \,\mathrm{fm}^{-3} = (0.56\,\mathrm{fm})^{-3}\,.
\label{rb0}
\end{align}
There is also a remnant branching density in the deconfined phase, but this 
is much smaller in space slices ($20\%$ of $\rho_B(0)$) than in time slices 
$(60\%$), in agreement with our geometrical discussion of vortex branching 
above. In fact, the branching density in time slices even increases slightly 
with the temperature within the deconfined phase. 

\begin{figure}[t]
	\begin{center}
		\includegraphics[width=8.7cm]{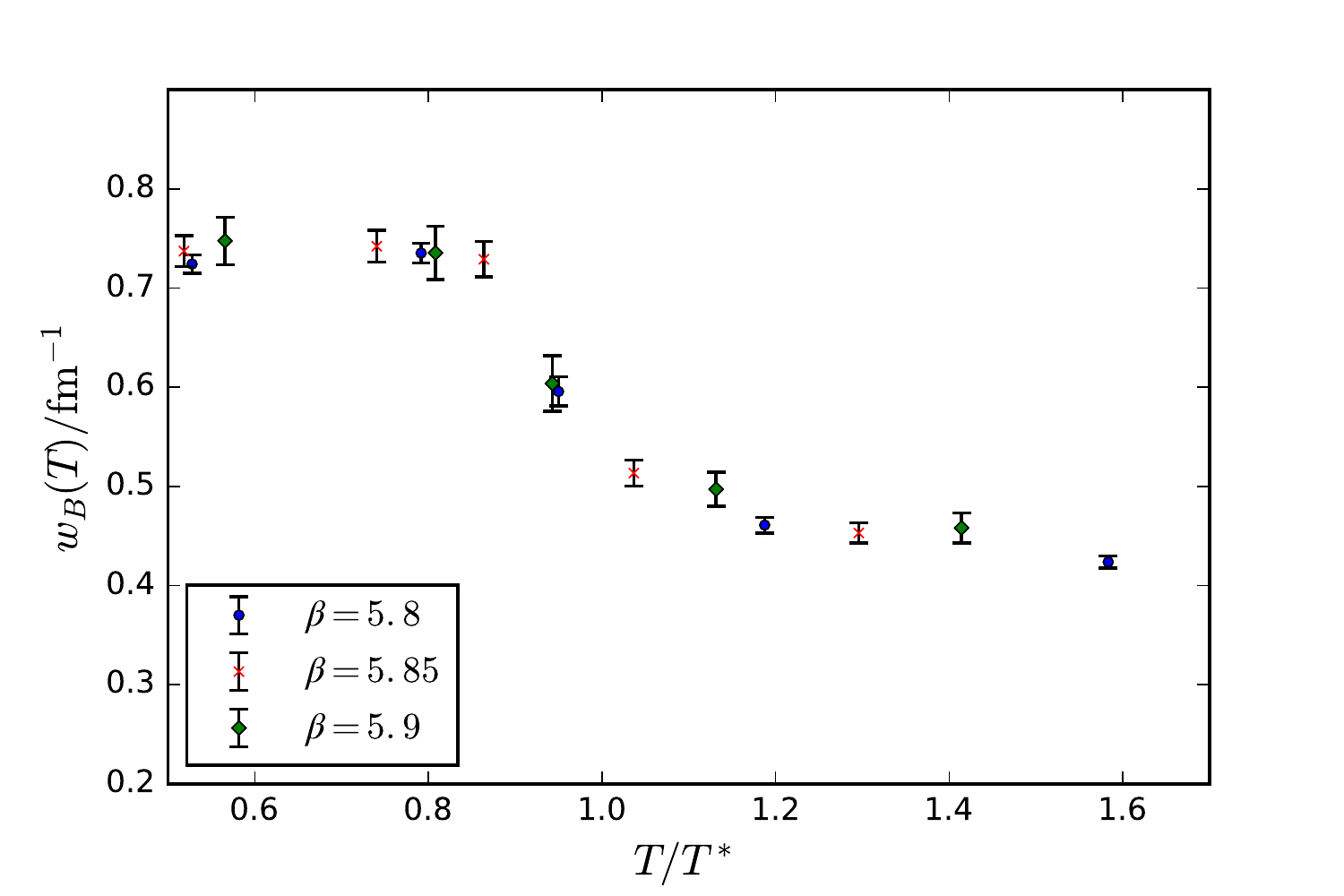}
		\includegraphics[width=8.7cm]{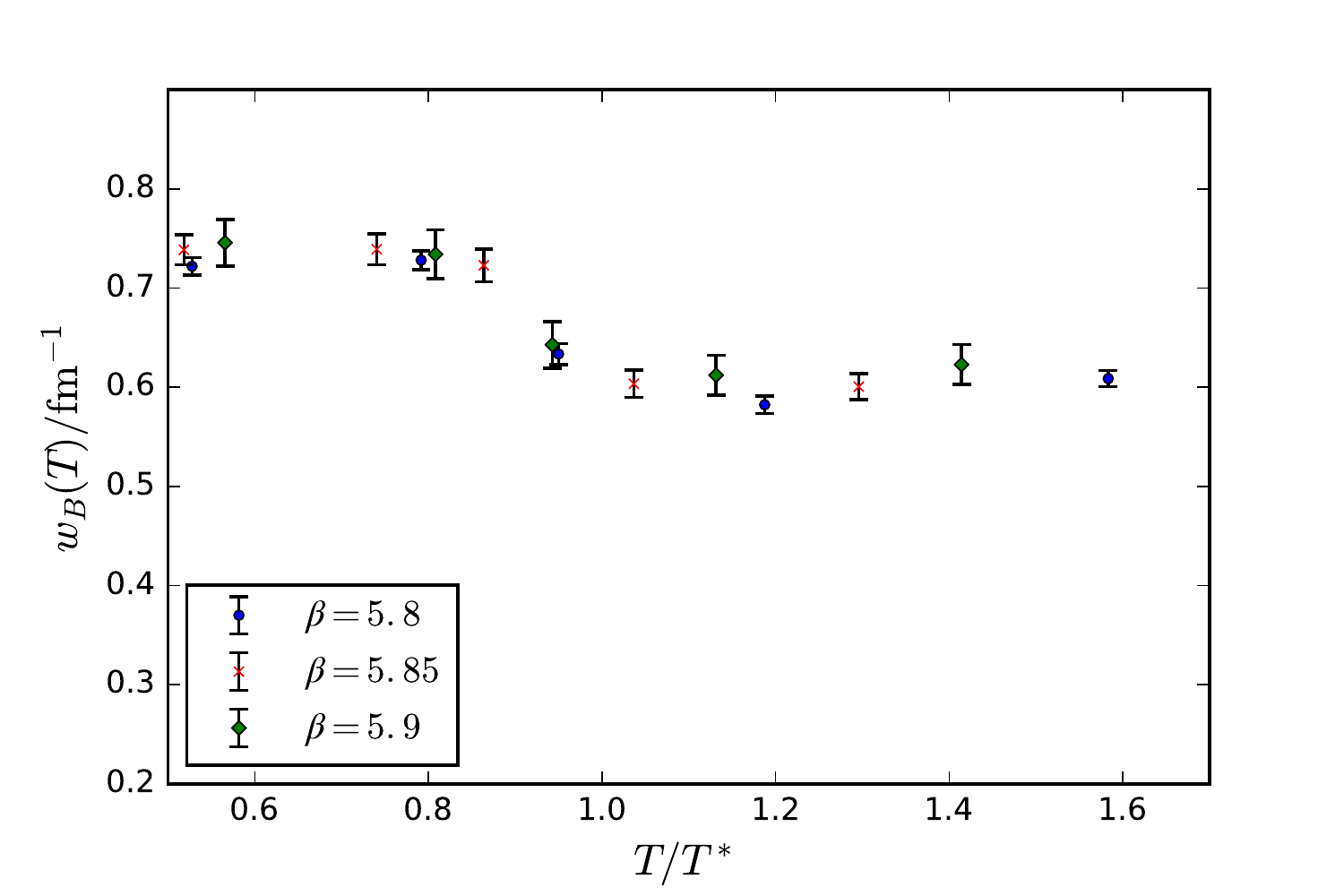}
	\end{center}
	\caption{Branching probability per unit length (\ref{wb}) in physical units, measured 
	in space slices (\emph{left}) and time slices (\emph{right}). Error bars include 
	statistical errors and uncertainties in the physical scale taken from Ref.~\cite{lucini}.}
	\label{fig:6}
\end{figure}

Next, we want to demonstrate that the steep drop in the branching density 
is \emph{not} due to an overall reduction of vortex matter itself, but 
rather signals a geometrical re-arrangement. Instead of studying $\rho_B / \rho$
directly, we make a small detour and first introduce the 
\emph{branching probability}
\begin{align}
q_B \equiv \frac{\text{\# elementary cubes in 3D slice with $\nu \in \{3,5\}$ }}
{\text{\# all elementary cubes in 3D slice with $\nu \neq 0$}}\,,
\label{qb}
\end{align}
which gives the likelihood that a vortex which enters an elementary cube of 
edge length equal to the lattice spacing $a(\beta)$ will actually branch 
within that cube. The branching probability $q_B$ itself cannot be a physical
quantity since it is expected to be proportional to the lattice spacing $a$ near the 
continuum limit.\footnote{To see this, assume that the probability of branching 
in a cube of edge length $a \ll 1$ is $q \ll 1$, and consider a cube of length $n a$
composed of $n^3$ sub-cubes of length $a$. Since vortices are stiff, most non-branching 
vortices do not change their direction if $a  \ll 1$ and just pass straight through $n$ 
sub-cubes. The probability of non-branching within the $n a$-cube is therefore 
$(1-q)^n$ at small spacing, so that the branching probability in the $n a$-cube 
becomes $1-(1-q)^n \approx n q$, i.e.~it is proportional to the edge length of the cube.}
This entails that the \emph{branching probability per unit length}
\begin{align}
w_B(T,\beta) \equiv \frac{q_B(T,\beta)}{a(\beta)}
\label{wb}
\end{align}
could be a physical quantity. As can be seen from Fig.~\ref{fig:6}, this is indeed the 
case as the curves for $w_B$ for all available couplings fall on a common curve. 
The temperature dependence of the physical quantity $w_B(T)$ is very similar to the 
branching density in Fig.~\ref{fig:4}, with the drop at $T=T^\ast$ being reduced from 
$75\%$ to about $50\%$. The qualitative features of the branching probability per unit 
length are, however, very similar to the branching point density, and both are physical 
quantities that scale to the continuum.

\begin{figure}[t]
	\begin{center}
		\includegraphics[width=12cm]{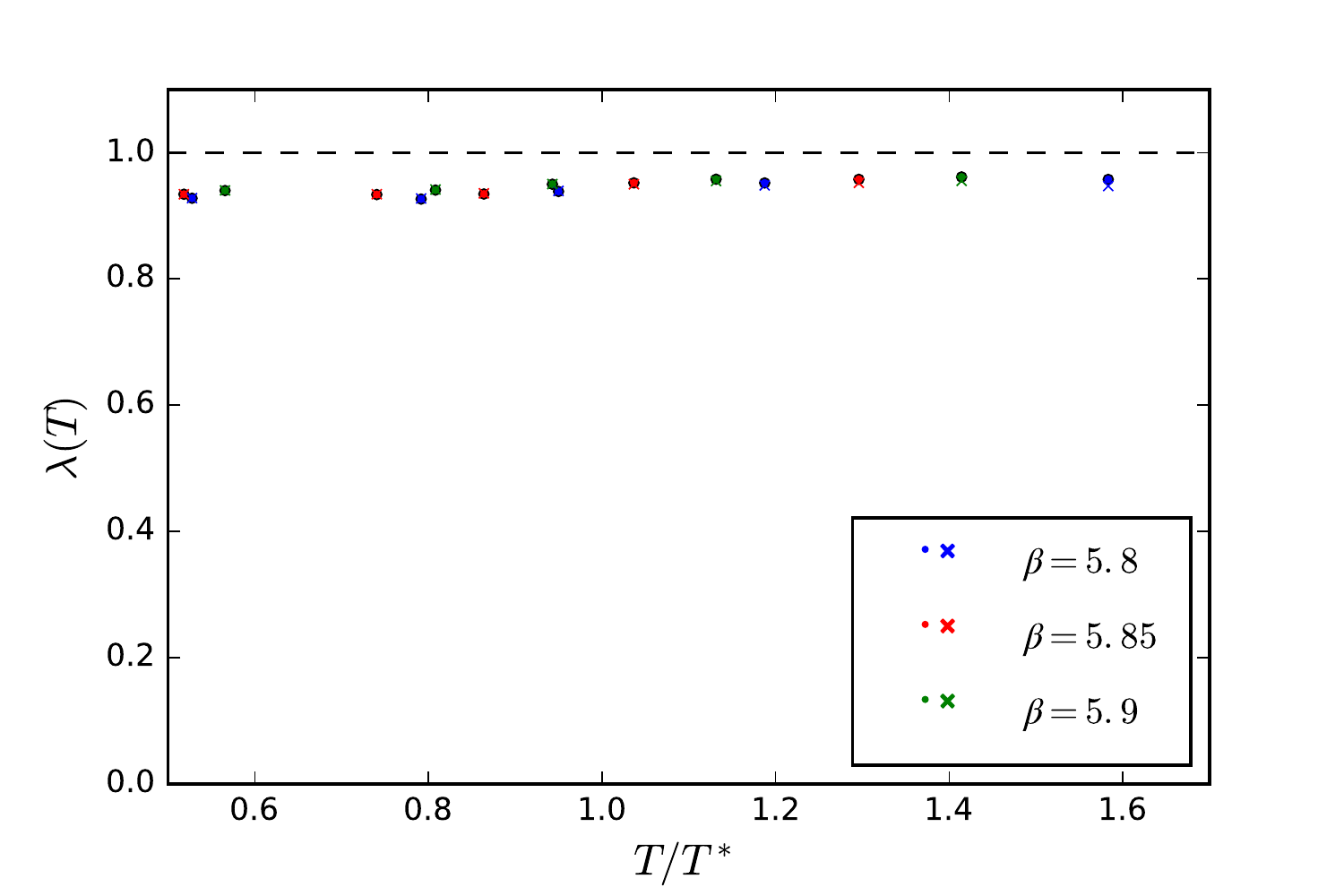}
	\end{center}
	\caption{The ratio $\lambda$ of physical quantities from eq.~(\ref{lambda}).
	Data comprises all available couplings and temperatures. Statistical errors 
	are generally at the $5\%$ level, but no error bars have been displayed to 
	improve the readability of the plot.}
	\label{fig:7}
\end{figure}

Next we want to show that the branching probability per unit length $w_B$ is actually 
related to the ratio $\rho_B / \rho$ of branching points and vortex matter density.
To see this, we consider an arbitrary 3D slice containing $V$ sites and thus also 
$V$ elementary cubes. The number of cubes of branching genus $\nu$ is denoted by $N_\nu$, 
and obviously $\sum_{\nu=0}^6 N_\nu = V$. Then the dimensionless branching density (\ref{rb}) 
can be expressed with eq.~(\ref{qb}) as 
\begin{align}
\hat{\rho}_B &= \frac{N_3 + N_5}{V} 
= q_B \cdot \frac{\sum\limits_{\nu=2}^6 N_\nu}{V}
= 3 q_B\,\frac{\sum\limits_{\nu=2}^6 \big[\nu + (2-\nu)\big] N_\nu}{6V}
= 3 q_B\,\frac{\sum\limits_{\nu=2}^6 \nu N_\nu}{6V}\cdot \left \{ 1 - 
\frac{\sum_{\nu=2}^6 (\nu-2) N_\nu}{\sum_{\nu=2}^6 \nu N_\nu}\right\}
= 3 q_B\,\hat{\rho}\,\lambda
\label{rel0}
\end{align}
with the dimensionless factor
\begin{align}
\lambda \equiv 1 - \frac{\sum\limits_{\nu=2}^6 (\nu-2) N_\nu}{\sum\limits_{\nu=2}^6 \nu N_\nu} 
\in [0,1]\,.
\label{lambda}
\end{align}
In the last step in eq.~(\ref{rel0}), we have used the fact that a cube with branching 
genus $\nu$ has $\nu$ non-trivial plaquettes on its surface, each of which is shared with an 
adjacent cube. Thus, the sum $\sum_\nu \nu N_\nu$ counts every non-trivial plaquette twice, 
and the dimensionless vortex area density eq.~(\ref{xvdens}) becomes, after averaging 
over all planes in the 3D slice,
\[
\hat{\rho} = \frac{\frac{1}{2}\,\sum\limits_{\nu=0}^6 \nu N_\nu}{3 V}
= \frac{\sum\limits_{\nu=2}^6 \nu N_\nu}{6V}\,,
\]
since a 3D slice with $V$ sites and periodic boundary conditions contains 
a total of $3V$ plaquettes.  After inserting appropriate factors of the lattice 
spacing in eq.~(\ref{rel0}), we obtain the exact relation
\begin{align}
\rho_B(T) = 3\,w_B(T)\,\rho(T) \,\lambda(T, a)\,.
\label{exa}
\end{align}
As indicated, the coefficient $\lambda$ may depend on the temperature and the 
lattice spacing, but it must fall in the range $[0,1]$. As a consequence, we obtain
an exact inequality between physical quantities,
\begin{align}
\rho_B(T) \le 3 \,w_b(T) \rho(T)\,,
\label{ineq}
\end{align}
which must be valid at all temperatures. Moreover, the deviation from unity in 
the coefficient $\lambda$ can be estimated, from eq.~(\ref{lambda}), 
\begin{align*}
\lambda = 1 - \frac{\sum\limits_{\nu=2}^6 (\nu-2)N_\nu}{\sum\limits_{\nu=2}^6 \nu N_\nu}
= 1 - \frac{N_3 + N_5}{\sum\limits_{\nu=2}^6 \nu N_\nu} + 
2 \,\frac{N_4 + N_5 + 2 N_6}{\sum\limits_{\nu=2}^6 \nu N_\nu} = 1 - 
\frac{1}{6}\,\frac{\hat{\rho}_B}{\hat{\rho}} + \mathcal{O}\big(\frac{N_4}{N_2}\big)
= 1 - \frac{1}{6}\,\frac{\rho_B(T)}{\rho(T)}\,a +  \mathcal{O}\big(\frac{N_4}{N_2}\big)\,.
\end{align*}
Here, the leading correction to unity vanishes in the continuum limit $a \to 0$ since both 
$\rho_B$ and $\rho$ are physical. Furthermore, the next-to-leading term has the simple 
branching $\nu=3$ removed and starts with the probability of self-intersection or osculation, 
which is small and presumably also proportional to $a$, by the same argument that led from 
eq.~(\ref{qb}) to eq.~(\ref{wb}) above. Thus, it is conceivable that 
$\lambda(T,a) = 1 + \mathcal{O}(a)$ and eq.~(\ref{exa}) turns into the relation
\begin{align}
w_B(T) = \frac{1}{3}\,\frac{\rho_B(T)}{\rho(T)}
\label{conject}
\end{align}
for $a \to 0$. This is renormalization group invariant. We have tested this conjecture 
numerically by computing the relevant coefficient $\lambda(T, a)$ from eq.~(\ref{lambda}).
The result is presented in Fig.~\ref{fig:7}, where we accumulate all available data for all 
temperatures and lattice spacings. As can be seen, $\lambda$ is indeed in the range $[0,1]$,
independent of temperature and very close to unity. Since the overall statistical uncertainty
is about $5\%$ and our calculations were all done at the lower end of the scaling window 
with a relatively large lattice spacing $a$, our numerics are at least compatible with 
$\lambda=1$ and hence eq.~(\ref{conject}) in the continuum limit. Further calculations with 
larger and finer lattices are clearly necessary to corroborate this conjecture.

Eq.~(\ref{conject}) shows that the drop of the branching density $\rho_B$ at the 
phase transition is \emph{not} due to an overall reduction of vortex matter $\rho$, 
since the branching probability per unit length, $w_B \sim \rho_B / \rho$ shows 
the same qualitative behaviour as $\rho_B$, even after scaling out the overall vortex
density. The conclusion is that both the branching point density
$\rho_B(T)$ from eq.~(\ref{rbx}) and the branching probability $w_B(T)$ per unit length 
eq.~(\ref{wb}) can be used as a reliable indicator for the phase transition,
and as a signal for the change in geometrical order of the vortices at the deconfinement 
transition. Our findings in full Yang-Mills theory match the  general expectations 
discussed above and also comply with the predictions made in the random vortex 
world-surface model \cite{engelhardt}.

\section{Conclusion}
In this work, we have studied the probability of center vortex branching within $SU(3)$ 
Yang-Mills theory on the lattice. The general expectation, confirmed only in models 
so far, was that the branching probability should be sensitive to the geometry of 
vortex clusters and thus provide an alternative indicator for the deconfinement
phase transition. We were able to corroborate this conjecture: both the branching 
point density $\rho_B(T)$ and the branching probability per unit length $w_B(T)$
are independent of the lattice spacing and exhibits a steep drop at the critical 
temperature, though a remnant branching probability remains even above $T^\ast$. 
This effect is much more pronounced in space slices of the original lattice, 
which clearly indicates a dominant alignment of vortices along the short time 
direction within the deconfined phase. The same conclusion can be drawn from 
the renormalization group invariant relation $w_B \sim \rho_B / \rho$, which 
proves that the drop in the branching density is \emph{not} due to an overall
reduction of the vortex matter $\rho$, but instead must be caused by the change 
in the geometry of the vortex cluster.

In future studies, it would be interesting to directly control the branching of 
vortices and study its effect on the confinement and the chiral symmetry breaking 
e.g.~through the Dirac spectrum in the background of such branching-free configurations.
The control over vortex branching could also address the obvious conjecture that the 
different (first) order of the phase transition for $G=SU(3)$ as compared to the weaker 
second order transition of $G=SU(2)$ is a result of the new geometrical feature 
of vortex branching.

\section*{Acknowledgment}
This work was supported by Deutsche Forschungsgemeinschaft (DFG) under 
contract Re 856/9-2.

\end{document}